\documentclass[apj]{emulateapj}
\usepackage{graphicx,amssymb,amsmath,epsfig}
\usepackage{natbib}

\def\apjl{ApJL}
\def\aj{AJ}
\def\apj{ApJ}
\def\nat{Nature}
\def\mnras{MNRAS}
\def\prd{Phys. Rev. D.}

\begin{document}
\normalsize

\author{Gilad Svirski and Ehud Nakar}
\affil{Raymond and Beverly Sackler School of Physics \&
Astronomy, Tel Aviv University, Tel Aviv 69978, Israel}

\title{SN 2008D: A Wolf-Rayet explosion through a thick wind}

\begin{abstract}
Supernova (SN) 2008D/XRT 080109 is considered to be the only direct detection of a shock breakout from a regular SN to date. While a breakout interpretation was favored by several papers, inconsistencies remain between the observations and current SN shock breakout theory. Most notably, the duration of the luminous X-ray pulse is considerably longer than expected for a spherical breakout through the surface of a type Ibc SN progenitor, and the X-ray radiation features, mainly its flat spectrum and its luminosity evolution, are enigmatic.
We apply a recently developed theoretical model for the observed radiation from a Wolf-Rayet SN exploding through a thick wind and show that it naturally explains all the observed features of SN 2008D X-ray emission, including the energetics, the spectrum and the detailed luminosity evolution. We find that the inferred progenitor and SN  parameters are typical for an exploding Wolf-Rayet. A comparison of the wind density found at the breakout radius to the density at much larger radii, as inferred by late radio observations, suggests an enhanced mass loss rate taking effect about ten days or less prior to the SN explosion. This finding joins accumulating evidence for a possible late phase in the stellar evolution of massive stars, involving vigorous mass loss a short time before the SN explosion.
\end{abstract}

\section{Introduction}\label{sec:intro}

%To this date, SN 2008D is the only direct observation of the electromagnetic birth moment of a standard SN, and the only time that the X-ray emission pattern that characterizes this phase was observed. While an X-ray pulse is indeed expected to announce the birth of a compact progenitor SN, the detailed features of the observed X-ray emission do not match a standard breakout through the progenitor surface. In this letter we argue that a SN explosion through a surrounding optically thick wind, a likely scenario for a Wolf-Rayet (WR) progenitor, offers an optimal explanation for the observations.

The X-ray transient XRT 080109, associated with SN 2008D, was discovered
by \emph{Swift}/XRT \citep{Soderberg+2008}. Later observations in Optical/UV led to its classification as a type Ibc SN, favoring a WR progenitor \citep{Soderberg+2008,Mazzali+2008,Malesani+2009,Modjaz+2009}.
The X-ray signal of SN 2008D is the most convincing candidate for a shock breakout of a standard SN. Unlike
other observed X-ray SN birth signals, which are all of rare broad line Ic SNe associated with Gamma-ray bursts \citep[e.g., SN2006aj/GRB060218;][]{Campana06},
the serendipitous discovery and the spectroscopic classification of SN 2008D suggest that it was a common signal, produced by a common Ibc SN \citep{Soderberg+2008}.

Nevertheless, the detailed X-ray observations are still unexplained. The initial rise-time, the following light curve and the spectrum are different than the ones predicted for a standard SN breakout, namely a spherical breakout from the stellar surface
\citep[e.g.][]{Chevalier+2008,Katz+2010,Nakar_Sari2010,Sapir+2011}.
In order to explain the prolonged rise-time, both a breakout through a thick wind \citep{Soderberg+2008} and an aspherical breakout \citep{Couch+2011} were suggested. However, lacking detailed theoretical models of the two scenarios (a  major progress
in modeling aspherical breakouts was only recently achieved, by \citealt{Matzner+2013}), neither could be confronted with the detailed observed spectrum and light curve. As often happens in the absence of a consensual explanation, a burst of a mildly relativistic jet was also suggested \citep{Xu+2008,Li+2008,Mazzali+2008}, although this model has no predictions in terms of the expected X-ray emission, that can be compared to the observations.

Recently we developed a detailed theoretical model for the emission from a SN breakout through a thick wind \citep{Svirski+2014}. Here we show, based on this model, that a scenario of a WR exploding through a thick wind naturally solves all inconsistencies, including the prolonged rise-time, and provides, without a need to invoke a significant breakout asphericity or an unconventional explosion scenario, an optimal explanation for the X-ray observations of SN 2008D. Moreover, the data fit all the model predictions, although these are tightly over-constrained, providing a strong support for this explanation.

In Section \ref{sec:observations} we describe the observations and indicate their tension with a standard SN breakout interpretation. We then summarize, in Section \ref{sec:model}, our theoretical model for WR SNe exploding through a thick wind \citep{Svirski+2012,Svirski+2014} and show, in Section \ref{sec:impl}, that this scenario explains the X-ray observations.
%We use our results to derive the physical properties of the pre-explosion wind and the radius and velocity of the shock breakout.
We conclude in Section \ref{sec:conclusion}.

\section{Observations vs. stellar surface breakout}\label{sec:observations}

\subsection{Observations}

Several groups analyzed the \emph{Swift} and \emph{Chandra} X-ray data \citep{Soderberg+2008,Li+2008,Mazzali+2008,Xu+2008,Modjaz+2009}, with results that are generally in good agreement across the groups. The analysis by \cite{Modjaz+2009} is the most comprehensive, and unless otherwise stated the numbers we quote refer to this work.

\begin{figure*}
	\centering
        \epsscale{1.1} \plotone{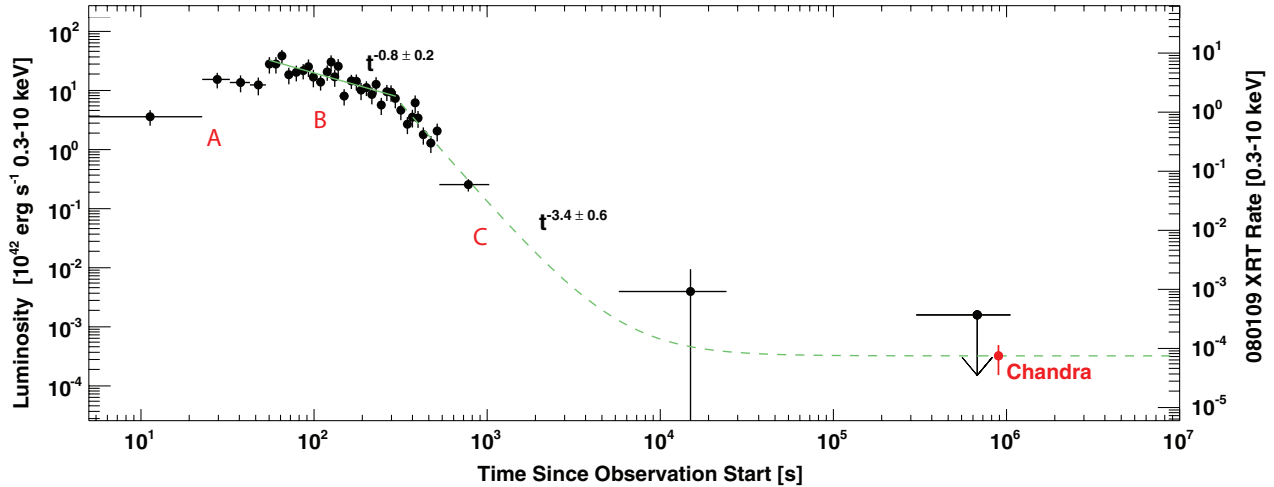}
	\caption{The X-ray luminosity evolution, adapted with permission from \cite{Modjaz+2009}. Black dots are \emph{Swift}/XRT data and the red dot is a \emph{Chandra} measurement. This light curve is expected, according to the model of \cite{Svirski+2014}, where (A) is the rising breakout pulse, (B) is the collisionless shock fast cooling phase, accompanied by a flat X-ray spectrum, and (C) is the slow cooling phase, with a fast decay of the X-ray signal.}
	\label{fig:evolution}
\end{figure*}

The rise-time of the X-ray transient is $50 \pm 30$ s according to \cite{Modjaz+2009} and $\sim 100$ s according to \cite{Soderberg+2008}, and we therefore take the rise-time to be $\approx 80$ s, consistent with both. The X-ray luminosity peaks at $3.8 \pm 1 \times 10^{43}\,\rm{erg/s}$, and \cite{Modjaz+2009} fits its evolution to a broken power-law in time: The luminosity at $60<t<300$ s ($t=0$ is the time of onset of the X-ray emission) is best described by a decaying power-law, $L_X \propto t^{\alpha}$, with an index $\alpha=-0.8 \pm 0.2$, whereas the luminosity at $t>300$ drops sharply, $\alpha=-3.4 \pm 0.6$.
The total estimated energy output in X-ray is $\sim 10^{46}$ erg \citep{Soderberg+2008,Chevalier+2008,Modjaz+2009}.
A \emph{Chandra} observation at day 10 reads $3.2 \pm 1.7 \times 10^{38}\,\rm{erg/s}$, indicating that the steep $t>300$ decay is halted around $10^4$ s or changes trend prior to the \emph{Chandra} observation. Figure \ref{fig:evolution} sketches the X-ray luminosity evolution during the first 10 days, based on Figure 1 from \cite{Modjaz+2009}.

All groups carried out a spectral fit of the X-ray radiation to a power-law model, $N(E)\propto E^{-\Gamma}$, integrating over the transient duration ($520$ s in \citealt{Modjaz+2009} and slightly shorter elsewhere). We quote here the results of these power-law fits and skip additional fits that involved a blackbody model, because a blackbody radiation is not expected according to our breakout interpretation.
The spectrum is best fit by a single power-law with a photon spectral index $\Gamma=2.3 \pm 0.3$ \citep{Soderberg+2008,Li+2008,Xu+2008,Mazzali+2008} or $\Gamma=2.1^{+0.3}_{-0.4}$ \citep{Modjaz+2009}. The power-law fit, across all groups, is consistent with a flat spectrum, $\nu F_{\nu}= Const$, or slightly softer, i.e. a comparable amount of energy in each photon logarithmic frequency scale across the frequency range of \emph{Swift}, $0.3-10$ keV. \cite{Soderberg+2008} also reported on a significant softening of the X-ray spectrum between the peak and the emission $400$ s later, during the rapid decay phase.

Later observations, at frequencies softer than X-rays, correspond to a standard Ibc SN \citep[e.g.][]{Soderberg+2008,Modjaz+2009}, and we only quote here the radio observations that are relevant for our analysis in Section \ref{sec:impl}.
\cite{Soderberg+2008} analyzed VLA observations, identified synchrotron emission, and inferred a shock radius $\approx 3\times 10^{15}$ cm at $t\approx 5$ d, implying a mean shock velocity $v \approx 0.25c$ (where c is the speed of light) over the first five days. In addition, they found that the observations indicate on a standard wind density profile, and assuming a wind velocity $v_w=1000$ km/s, they inferred for the above radius a mass loss rate $\dot{M}\approx 10^{-5}\rm{M_{\odot}\,yr^{-1}}$.

\subsection{A standard breakout?}

In a standard breakout scenario, namely a spherical shock breakout through a stellar surface, the duration of the initial pulse is dominated by the light crossing time across the progenitor \citep[e.g.][]{Klein_Chevalier1978}. The radius of a WR is expected to reach up to a few times $10^{11}$ cm, implying a light crossing time of order $10$ s, in contrast with the observed $\sim 300$ s of $L_X \ge 10^{43}\,\rm{erg/s}$. We note that the progenitor radius is under debate, ranging (based on UV/Opt observations) from $10^{11}$ cm \citep{Soderberg+2008,Rabinak+2011} to $10^{12}$ cm \citep{Chevalier+2008}, but even the latter estimate matches a light crossing time that is an order of magnitude below the observed duration.

One could argue that the moderate $\alpha=-0.8 \pm 0.2$ initial luminosity decay, lasting until $t\approx 300$ s, is marginally consistent with the $\alpha=-4/3$ expected during the planar phase of a standard breakout \citep[][hereafter NS10]{Piro+2010,Nakar_Sari2010}. However, a planar time $R_*/v\sim 300$ s implies, for $v=0.25c$ \citep{Soderberg+2008}, a stellar radius $R_*\sim 4\times 10^{12}$ cm. These $R_*$ and $v$ correspond to a breakout energy $\sim 10^{47}$ erg, an order of magnitude above the observed, and a UV signal that is too bright \citep{Chevalier+2008}. While uncertainty in the rise-time and the breakout velocity may reduce this tension, it cannot completely remove it. In addition, such a radius is not expected for a WR progenitor, as implied by the Ibc classification.

Regarding the spectrum, the observed radiation from a WR SN breakout is expected to deviate significantly from thermal equilibrium (\citealt{Katz+2010}; NS10), and therefore a blackbody spectrum is not expected. Instead, we expect a radiation that peaks at a few keV, likely within the \emph{Swift}/XRT $0.3-10$ keV detection window (e.g. NS10; \citealt{Sapir+2011}), but is fainter elsewhere. NS10 predict a $\nu F_{\nu}\propto \nu^{\beta}$ with $0.5<\beta<1$ up to the peak and an exponential decay above it.
%As \cite{Sapir+2011} mention, in a standard breakout scenario we expect a positive spectral slope (in $\nu F_{\nu}$) for the low energy tail regardless of the assumed instantaneous spectrum, as both the luminosity and the radiation temperature drop with time.
However, when fitted to a power-law spectrum, the observations are consistent with a flat or slightly negative spectrum across the complete \emph{Swift} detection range.

\section{The emission from a Wolf-Rayet SN exploding through a thick wind}\label{sec:model}

SN ejecta that expand into a surrounding wind give rise to an interaction layer, composed of a forward shocked wind and a reverse shocked ejecta. Assuming a spherical shock, an outer (pre-explosion) stellar envelope density profile with a polytropic index of $3$, and a standard wind density profile, $\rho_w\propto r^{-2}$, the radius and velocity of the interaction layer evolve as $r\propto t^{0.875}$ and $v\propto t^{-0.125}$, respectively \citep{Chevalier1982}. The optical depth of the unshocked wind ahead of the shock evolves as $\tau\propto r^{-1}$, or approximately $\tau\propto t^{-1}$

In \cite{Svirski+2014} we derive a model for the radiation from a Wolf-Rayet SN exploding through a thick wind, and below we quote the relevant findings. If the optical depth of the wind surrounding the star is $>c/v$, where $c$ is the speed of light and $v$ is the SN shock speed, then the radiation mediated shock that crosses the star envelope continues into the wind, and breaks out when the optical depth of the unshocked wind decreases to $\tau\approx c/v$. The bolometric luminosity while $\tau>1$ is:
\begin{equation}\label{eq:L(t)}
L(t)\approx 3.5 \times 10^{43}\,t_{bo,m}\,v_{bo,10}^3 \left(\frac{t}{t_{bo}}\right)^{-0.4}\,\rm{\frac{erg}{s}}
\end{equation}
where $v_{bo}$ is the shock velocity at the time of breakout,
$v_{bo,10}=v_{bo}/10^{10} {\rm~cm~s^{-1}}$, $t_{bo}$ is the breakout
time and $t_{bo,m}=t_{bo}/{\rm minute}$.
Observationally, $t_{bo}$ is
also roughly the rise time of the breakout pulse.

The shock velocity at the breakout depends on the explosion energy, $E$, the SN ejecta mass, $M_{ej}$, and the breakout time:
\begin{equation}\label{eq:v_bo}
    v_{bo} \approx 6\times 10^9 {\rm~cm/s~} M_{5}^{-0.31} E_{51}^{0.44}
    t_{bo,m}^{-0.25},
\end{equation}
where $E_{51}=E/10^{51}$ erg and $M_{5}=M_{ej}/5M_\odot$.

The breakout pulse has a non-thermal spectrum that peaks at a few keV. Unlike a standard breakout, the rise-time is $\sim \frac{R_{bo}}{v_{bo}}$ (where $R_{bo}$ is the breakout radius) rather than $\sim \frac{R_*}{c}$, and the shock breakout has no planar phase. Following the breakout, a collisionless shock replaces the radiation mediated shock, and a layer of hot shocked electrons, with a temperature $T_h\ge 60$ keV (hereafter, a temperature $T$ denotes an energy $k_BT$, where $k_B$ is Boltzmann constant), forms behind the shock \citep{Katz+2011,Murase+2011}. The efficient cooling of these hot electrons dominates the SN luminosity, and is sustained by inverse Compton over soft photons that were deposited by the radiation mediated shock that crossed the star. This soft radiation injects into the interaction region a fraction $f_{inj}\approx \frac{1}{5} \frac{R_*}{R_{bo}}$ of the luminosity powered by the interaction, at a characteristic temperature $T_{inj}$. The condition
\begin{equation}\label{eq:rough}
\frac{T_{inj}}{m_ec^2/\tau^2} \ll f_{inj} \ll 1,
\end{equation}
where $m_e$ is the electron mass, implies a flat or nearly flat spectrum, $\nu F_{\nu}\approx Const$ across the range $T_{inj}\lesssim T\lesssim \min(m_ec^2/\tau^2, T_h)$. This condition is typically satisfied for WR stars exploding through a thick wind.

At $\tau\ll 1$, the emission takes the form of a standard core-collapse SN, with no wind. The interaction signal is fainter, and dominated by single scattering of soft photons by the hot shocked layer. However, a small fraction of the soft photons goes through multiple scattering and produces an X-ray signal. Since the temperature of the hot shocked layer at this stage is $\sim 200$ keV, each collision with such electron upscatters a photon by an average factor $\sim 4$, and it takes $n=\ln(T_X/T_{SN})/\ln(4)$ collisions to bring a soft $T_{SN}$ photon to an X-ray temperature $T_X$. A signature of such X-ray signal is a decay pattern $L_X(t)\propto t^{-n}$, and a continuous softening of the X-ray spectrum.

\section{A coherent picture of SN 2008D}\label{sec:impl}

Our model predicts the X-ray luminosity evolution and the spectrum observed from a SN exploding through a standard WR wind. Assuming such a scenario, the luminosity evolution depends on $t_{bo}$, which is directly observed, and $v_{bo}$, which can be inferred from the observations in multiple independent ways. The spectrum at $\tau\gtrsim 1$ depends on $f_{inj}$ and $T_{inj}$, both inferred from theoretical considerations.
Below we show that the predictions of our model, despite being over-constrained in several ways by the observations, are all satisfied.

The most robust constraint on the shock breakout velocity comes from the observed luminosity peak, which value is highly sensitive to the velocity but far less so to the breakout time, and rather insensitive to the time since breakout (see Equation \ref{eq:L(t)}). Substituting $t_{bo}=80$ s and $L_{bo}=3.8 \pm 1 \times 10^{43}\,\rm{erg/s}$ \citep{Modjaz+2009} in Equation \ref{eq:L(t)} yields $0.84<v_{bo,10}<1$.
An independent alternative estimate of $v_{bo}$ comes from the optical depth at breakout:
Following the breakout, due to fast cooling, the X-ray luminosity decays slowly as the shock traverses the range $1<\tau<\tau_{bo}$. As $\tau\propto t^{-1}$, the ratio between the time of transition to a sharp luminosity drop and the rise time indicates the breakout optical depth, $\tau_{bo}\approx\frac{300}{80}\approx 4$, implying $v_{bo,10}\approx 0.8$, in good agreement with the first estimate.
Yet a third independent estimate is obtained from Equation \ref{eq:v_bo}. Combining a progenitor mass $5M_{\odot}\lesssim M \lesssim 7M_{\odot}$ and an explosion energy $2\lesssim E_{51}\lesssim6$ \citep{Soderberg+2008,Mazzali+2008} with $t_{bo}=80$ s implies $0.75\lesssim v_{bo,10}\lesssim 1.3$, providing a second sanity check.
Combining these three estimates and accounting for the uncertainty in the rise-time, the peak luminosity, and the exact $\tau$ of transition to a slow cooling, we infer a breakout velocity in the range $0.6\lesssim v_{bo,10}\lesssim 1.2$.

We next examine the observed spectrum of the X-ray transient. The X-ray radiation is dominated by the cooling of the collisionless shock that forms after the breakout, and its spectrum is determined by $f_{inj}$ and $T_{inj}$. We estimate $T_{inj}$ to be the characteristic SN radiation temperature calculated in NS10 Equation 41. The published range of the progenitor mass and explosion energy, along with an assumed WR radius $R_*=5R_{\odot}=3.5\times 10^{11}$ cm, imply an initial $T_{inj}\sim 300$ eV at $t=80$ s, dropping sharply to $T_{inj}\sim 30$ eV at $t=300$ s. A simple estimate $v_{bo}t_{bo}$ implies a breakout radius $R_{bo}\approx 6\times 10^{11}$ cm, and a respective $f_{inj}\sim 0.1$ for $R_*= 5R_{\odot}$.
These $f_{inj}$ and $T_{inj}$ values satisfy Equation \ref{eq:rough} and imply, at $80\lesssim t \lesssim 300$ s, a flat spectrum that initially spans across the range $0.3\lesssim T \lesssim 50$ keV, and later widens to $0.03\lesssim T \lesssim 100$ keV. This is consistent with the flat spectrum observed by \emph{Swift}/XRT within its detection window of $0.3-10$ keV. Note that while a radius smaller than $5R_{\odot}$ by a factor of a few implies somewhat lower $f_{inj}$ and initial $T_{inj}$, these still yield a rather flat spectral slope. Hence, an approximately flat spectrum is a robust prediction of our model and it is unaffected by the freedom, in the progenitor mass and radius and in the explosion energy, implied by current range in these values presented in the literature.

The flat X-ray spectrum also explains the first phase of the X-ray luminosity evolution. Inverse Compton provides a fast cooling of the shock as long as $\tau\gtrsim 1$, and therefore one would expect Equation \ref{eq:L(t)} to apply for $80\lesssim t\lesssim 300$ s. However, the observations indicate that over this period $L_X \propto t^{\alpha}$, with an index $\alpha=-0.8 \pm 0.2$, only marginally consistent with the predicted $\alpha=-0.4$ implied by Equation \ref{eq:L(t)}. In fact, this difference is expected: While most of the initial X-ray spectrum is covered by the \emph{Swift}/XRT window, by $t=4t_{bo}$ about half of the spectral logarithmic range falls outside the \emph{Swift}/XRT window, implying an X-ray decay that is slightly faster than that predicted for the bolometric luminosity, as observed.

At $\tau<1$ ($t>300$ s) the shock becomes slow cooling and the soft SN radiation soon becomes the dominant component of the bolometric luminosity. At this stage, an X-ray signal decaying as $t^{-n}$ is expected. An initial photon temperature of a few eV, matching the first few hours after the explosion, implies that three subsequent Compton scatters are required to upscatter a photon to within the \emph{Swift}/XRT detection window. Therefore the X-ray luminosity observed by \emph{Swift}/XRT should follow $L_X \propto \tau^3 \propto t^{-3}$, in agreement with the observed X-ray luminosity at $t>300$ s (after a few hours $n$ should rise to four, still consistent with the observed slope). Here $\tau$ is the optical depth of the hot layer, which is comparable, due to the slow cooling, to the optical depth of the unshocked wind. The X-ray signal at this stage is expected to soften, as indeed reported by \cite{Soderberg+2008}.

At day $10$ \emph{Chandra} recorded an X-ray luminosity $3.2 \pm 1.7 \times 10^{38}\,\rm{erg/s}$ \citep{Modjaz+2009}, higher than expected by the $t^{-3}$ trend described above. This may be accounted for by inverse Compton upscattering of Optical photons to X-ray by relativistic electrons, accelerated by the collisionless shock. \cite{Soderberg+2008} reported observation of synchrotron emission, an imprint of relativistic electrons, and estimated, based on \cite{Waxman+2007}, an expected X-ray luminosity of $\sim 5 \times 10^{38}\,\rm{erg/s}$, in agreement with \emph{Chandra} observation. \cite{Chevalier+2008} further supported this late X-ray origin.

Applying our model to SN 2008D suggests an increased mass loss rate during the days that preceded the explosion: The mass loss rate inferred by \cite{Soderberg+2008}, $\dot{M}\approx 10^{-5}\rm{M_{\odot}\,yr^{-1}}$, assuming $v_w=1000$ km/s, corresponds, at $r=3\times 10^{15}$ cm, to the rate a year before the explosion, and it is consistent with the average values inferred for Galactic WR stars \citep[e.g.][]{Cappa+2004}. However, if $v_w=1000$ km/s is assumed, a $\tau_{bo}\approx 4$ at $R_{bo}\approx 6\times10^{11}$ cm requires a much higher mass loss rate shortly before the explosion, $\dot{M}\approx 2\times 10^{-4}\rm{M_{\odot}\,yr^{-1}}$, equivalent to a wind density $\sim 10^{13}\,\rm{cm^{-3}}$ at $R_{bo}$. These values are on the high end of those observed in galactic WR stars. We can use \cite{Soderberg+2008} mean velocity over the first five days, $\approx 7.5\times 10^9 \,\rm{cm\,s^{-1}}$, to estimate the duration of the enhanced mass loss episode. While a standard wind density profile extending to $3\times 10^{15}$ cm implies (due to the $v\propto t^{-0.125}$ deceleration) $v_{bo}\approx 1.5\times 10^{10} \,\rm{cm\,s^{-1}}$, higher than the $v_{bo}$ inferred by our model, a termination of the enhanced wind density profile at $\lesssim 10^{14}$ cm, matching an enhanced mass loss starting up to about ten days before the explosion, can accommodate our inferred $v_{bo}\lesssim 1.2\times 10^{10} \,\rm{cm\,s^{-1}}$.

\section{Conclusion}\label{sec:conclusion}

We offer a first coherent picture of SN 2008D X-ray observations, including a first explanation for the observed flat X-ray spectrum across the \emph{Swift}/XRT $0.3-10$ keV window during the first few hundred seconds, and a first account for the complete X-ray luminosity evolution of SN 2008D/XRT 080109, from breakout to $t\sim 1$ d.
We find that a typical WR progenitor and typical SN parameters offer the optimal explanation for the observations.

We apply our model for WR SNe exploding through a thick wind \citep{Svirski+2014} to the X-ray observations of SN 2008D. The model allows only little freedom in deriving physical parameters from the observations, and it over-constrains the shock velocity. In addition, it enforces a flat X-ray spectrum at $\tau \gtrsim 1$ and a $t^{-3}$ to $t^{-4}$ X-ray luminosity decay at $\tau\ll 1$. Hence, the combination of an observed $t_{bo}$ and an observationally inferred $v_{bo}$ uniquely determines the X-ray luminosity and spectrum as a function of time. Despite these tight constraints, the observations satisfy all the predictions of the model.
%, thus providing multiple independent lines of evidence that support a breakout through wind scenario.
Accordingly, SN 2008D had a compact progenitor ($R_*\lesssim3\times10^{11}$ cm), presumably a WR, that exploded through a thick wind of a standard wind density profile, $\rho\propto r^{-2}$, with $R_{bo}\approx 6\times10^{11}$ cm, $v_{bo}\approx 8\times 10^9 \,\rm{cm\,s^{-1}}$, $\tau_{bo}\approx 4$ and $n_{bo}\sim 10^{13}\,\rm{cm^{-3}}$.

Observations of SN 2008D suggest that the explosion of the core was aspherical \citep{Maund+2009,Gorosabel+2010}. According to a recent theoretical prediction, if such explosion gives rise to a significant deviation from spherical symmetry of the shock at the breakout, it implies a breakout shock velocity that is lower than a spherical breakout velocity, over major parts of the shock front \citep{Matzner+2013}. The good agreement between our spherical model predictions and the observations, and the fact that the breakout velocity we infer, as well as the one estimated by \cite{Soderberg+2008}, are both rather high, as expected for spherical WR shock breakouts, suggest that the asphericity level of the shock at the breakout was low, and its effect on the X-ray emission, as well as the dynamics, was minor.

Interestingly, a comparison of the shock velocity and wind density that we find at $R\sim10^{12}$ cm, to that inferred by radio observations at $R\sim3\times10^{15}$ cm, implies that the mass loss rate of the progenitor has increased by more than an order of magnitude during the few days that preceded the SN explosion.
A pre-explosion enhanced mass loss may be a common feature of massive stars. \cite{Ofek+2013} reported a mass-loss event just a month before SN 2010mc, likely the explosion of a luminous blue variable progenitor, and suggested a causal connection between the pre-explosion mass loss burst and the explosion. Based on a sample of 16 type IIn SNe, \cite{Ofek+2014} inferred that at least half of the type IIn SNe experience a similar pre-explosion outburst. \cite{Gal-Yam+2014} identified SN 2013cu as an explosion of a WR progenitor and found indications for an increased mass loss rate starting a year before the explosion. Our results for SN 2008D join these accumulating indications and suggest that at least some WR progenitors experience an increased mass loss rate during a short period prior to their explosion. The excess mass loss in this case is due to a higher continuous mass loss, rather than the pre-explosion bursts that may characterize many type IIn SNe. Put together, these findings may indicate on a general phase of enhanced mass loss in the late stellar evolution of massive stars, e.g., by a process as suggested by \cite{Shiode+2014}.

G.S. and E.N. were partially supported by an ISF grant (1277/13), an ERC starting
grant (GRB-SN 279369), and the I-CORE Program of the Planning and Budgeting Committee and The Israel Science Foundation (1829/12).

\end{document}